\documentclass[preprint,preprintnumbers,amsmath,amssymb,floatfix,nofootinbib]{revtex4}
\usepackage{epsf}
\usepackage{graphicx}
\usepackage{subfigure}
\begin{document}
\title{Threshold Resummed and Approximate NNLO results for $W^+W^-$ Pair Production at the LHC}
\author{S.~Dawson$^{a}$, Ian~M.~Lewis$^{a}$, and Mao~Zeng$^{b}$}

\affiliation{
$^a$Department of Physics,
Brookhaven National Laboratory\\
Upton, N.Y.~ 11973, USA\\
$^b$ C.~N.~Yang Institute for Theoretical Physics\\
Stony Brook University, Stony Brook, N.Y.~11794\\
\vspace*{.5in}}

\date{\today}

\begin{abstract}
The next-to-leading order (NLO)  QCD radiative corrections to $W^+W^-$ production at hadron colliders are well understood.  We 
combine NLO perturbative QCD calculations with soft-gluon resummation of threshold logarithms to find  a next-to-next-to
leading logarithmic (NNLL) prediction for the total cross section and the invariant mass distribution at the LHC.  We also obtain approximate next-to-next-to-leading order (NNLO) results for the total $W^+W^-$ cross section at the LHC which 
includes all contributions from the scale dependent leading singular terms. Our result for the approximate NNLO total cross section is the most precise theoretical prediction available.  Uncertainties due to scale variation are shown to be small
when the threshold logarithms are included. NNLL threshold resummation increases the $W^+W^-$ invariant mass distribution by $\sim 3-4\%$ in the peak region for both $\sqrt{S}=8$ and $14$ TeV.  The NNLL threshold resummed and approximate NNLO cross sections increase the NLO cross section by $0.5-3\%$ for $\sqrt{S}=7,\,8,\,13,$ and $14$ TeV.

\end{abstract}

\maketitle
\section{Introduction}
Exploring the Higgs and electroweak sector of the Standard Model is one of the primary goals of the LHC.  The pair production of gauge bosons is important both as a test of the $SU(2)\times U(1)$ gauge structure and as a background for Higgs boson searches.  Precise predictions for both total and differential cross sections are needed in order to understand the shape of the background to the Higgs signal and to search for anomalous three gauge boson couplings.  Understanding the theoretical prediction for $pp\rightarrow W^+W^-$ is particularly important for the measurement of the Higgs decay channel,   $H\rightarrow W^+W^-\rightarrow2\ell2\nu$, where there is no resonant structure. The $W^+W^-$ background is estimated by a sideband analysis, where the cross section is normalized via a control region with a minimum dilepton invariant mass.  Using Monte Carlo, the line shapes of the $W^+W^-$ distributions are then extrapolated into the Higgs signal region~\cite{CMS:bxa,ATLAS:2013wla}.  A change in the $W^+W^-$ invariant mass distribution will alter the dilepton invariant mass distribution, and consequently change the extrapolation of the background estimates in the Higgs signal region.

 The production of $W^+W^-$ pairs  with a subsequent leptonic decay has been studied at the Tevatron~\cite{Aaltonen:2009aa,Abazov:2009ys}, while both ATLAS~\cite{ATLAS:2012mec} and CMS~\cite{Chatrchyan:2013oev,Chatrchyan:2013yaa} have reported results at the LHC.  The LHC results for the total $W^+W^-$ cross section,
\begin{eqnarray}
\hbox{ATLAS},~\sqrt{S}=7~{\rm TeV}~~&\sigma=&51.9\pm
2.0
(\hbox{stat})\pm 3.9
(\hbox{syst})\pm 2.0
(\hbox{lumi})~\hbox{pb}
\nonumber \\
{\hbox{CMS}},~\sqrt{S}=7~{\rm TeV}~~&\sigma=&52.4\pm 2.0
(\hbox{stat})\pm 4.5
(\hbox{syst})\pm 1.2
(\hbox{lumi})~
\hbox{pb}
\nonumber \\
{\hbox{CMS}},~\sqrt{S}=8~{\rm TeV}~~&\sigma=&69.9\pm 2.8
(\hbox{stat})\pm 5.6
(\hbox{syst})\pm 3.1
(\hbox{lumi})~
\hbox{pb}\, ,
\end{eqnarray}
are slightly higher than the Standard Model predictions at next-to-leading order (NLO) in QCD~\cite{Campbell:2011bn},\footnote{The theoretical predictions have been evaluated at NLO using MCFM with MSTW2008nlo PDFs and a central scale
choice of $\mu_f=M_W$.  The uncertainties shown in Eq.~(\ref{theorypred}) result from varying the scale up and down by a factor of $2$.  The predictions of Eq.~(\ref{theorypred}) include the next-to-next-to-leading order (NNLO) contribution from the $gg$ initial state~\cite{Campbell:2011bn}.}
\begin{eqnarray}
&\sigma(\sqrt{S}=7~{\rm TeV})_{\hbox{Theory}}&=
47.04^{+2.02}_{-1.51}~\hbox{pb}
\nonumber \\
&\sigma(\sqrt{S}=8~{\rm TeV})_{\hbox{Theory}}&=57.25^{+2.347}_{-1.60}~\hbox{pb}\, .
\label{theorypred}
\end{eqnarray}
The slight differences between the measured LHC values and the MCFM NLO predictions have  led to speculation that the  measured $W^+W^-$ cross section is a subtle sign of new physics~\cite{Feigl:2012df,Curtin:2012nn,Curtin:2013gta}.  

The NLO QCD corrections to $pp\rightarrow W^+W^-$ were computed in Refs. \cite{Frixione:1993yp,Ohnemus:1991kk}, and then extended to include leptonic decays in Ref. \cite{Dixon:1998py}.   The NLO predictions for the total cross section have a $3-4\%$ uncertainty at the LHC
 due to the choice of parton distribution functions (PDFs) and renormalization/factorization scale variation~\cite{Campbell:2011bn}.  The contribution from $gg\rightarrow W^+W^-$ is formally NNLO, but numerically contributes $\sim 3\%$ at $\sqrt{S}=7$~TeV and $\sim 4\%$ at $\sqrt{S}=14$~TeV~\cite{Dicus:1987dj,Glover:1988fe,Binoth:2006mf,Binoth:2005ua}.  The NLO results have been interfaced with a shower Monte Carlo using the formalism of the POWHEG box~\cite{Melia:2011tj,Hamilton:2010mb,Hoche:2010pf}.  The electroweak corrections and the contribution from the $\gamma\gamma$ initial state are also known and contribute less than $1-2\%$ to the total cross section at the LHC~\cite{Bierweiler:2012kw,Baglio:2013toa}.  These corrections are enhanced at large values of the $W^+W^-$ invariant mass, but have opposite signs and largely cancel.

 In this paper, we extend these results by including a resummation of threshold logarithms in  the prediction of $W^+W^-$ production. Previously, the resummation of large logarithms associated with gluon emission at low transverse momentum, $p_T$, in $W^+W^-$ production was considered~\cite{Grazzini:2005vw}.  Unlike $p_T$ resummation which is normalized to the NLO cross section, the emission of soft gluons near threshold  can potentially enhance the rate.  We consider the next-to-leading logarithmic (NLL) and next-to-next-to-leading logarithmic (NNLL) resummation of threshold corrections.  To accomplish this, we utilize the formalism of soft collinear effective theory (SCET)~\cite{Bauer:2000ew,Bauer:2000yr,Bauer:2001yt,Beneke:2002ph} which allows the resummation to be performed directly in momentum space~\cite{Becher:2006mr,Becher:2006nr}. This formalism has been used for processes with colorless final states such as Drell-Yan~\cite{Becher:2007ty}, Higgs production~\cite{Ahrens:2008qu,Ahrens:2008nc,Ahrens:2010rs},  associated W/Z plus Higgs production~\cite{Dawson:2012gs}, direct photon production~\cite{Becher:2009th} and SUSY slepton pair production~\cite{Broggio:2011bd}.  
 
The SCET formalism has also been applied to top quark pair production to resum the threshold corrections to the invariant mass distribution and to the total cross section~\cite{Ahrens:2010zv,Ahrens:2011mw,Ahrens:2011px}.   The total cross section for top quark pair production using threshold resummation has been obtained using two different sets of
threshold limits:  one starting from the invariant mass distribution of the top quark pair (pair invariant mass kinematics) and the other beginning from the transverse momentum or 
rapidity distribution of the top quark\cite{Ahrens:2010zv,Ahrens:2011mw,Kidonakis:2010dk}.  
The total cross section is then obtained by integrating over the resummed distributions.  Within the theoretical uncertainties, the total cross sections 
for top quark pair production obtained with the different starting points are in reasonable agreement\cite{Ahrens:2011mw,Ahrens:2011px}.  
The total cross section can also be obtained in the threshold limit,
$\beta\rightarrow 0$, where $\beta$ is the top quark velocity, and the terms of ${\cal O}(\alpha_s^n\ln^m\beta)$ 
resummed\cite{Beneke:2009ye}.
  At the LHC and the Tevatron, however, the
largest contributions to the total cross section for top quark pair production
 are not from the $\beta\rightarrow 0$ region.  In this work, we will use  pair invariant mass kinematics for the $W^+W^-$
final state to obtain our resummed results. 
 
 The calculation of differential cross sections involves several scales.  We consider the threshold limit $z\equiv {M_{WW}^2\over {s}}
 \rightarrow 1$ which dynamically becomes important due to the fast decline of the parton luminosity function as $\tau$ increases \cite{Becher:2007ty}, with $M_{WW}$ the invariant mass of the $W^+W^-$ pair and ${s}$ the partonic center of mass energy. Near the partonic threshold, up to subleading powers of $(1-z)$, the cross section factorizes into a soft function which describes the soft gluon emissions and a hard function which includes the virtual corrections to the cross section. We can combine the NLO soft and hard functions with their renormalization group (RG) evolution equations to give NNLL resummed results which resum large logarithms of the form  $\alpha_s^n\biggl({\ln^m(1-z)\over 1-z}\biggr)_+$ with $ m \leq 2n-1$. Alternatively, the RG evolution of the hard function, known to NNLO, can be matched with exact NLO results for the hard function to obtain the approximate NNLO hard function which includes the leading scale dependent contributions. Combined with the known NNLO soft function \cite{Becher:2007ty} for color singlet production, we are able to obtain the approximate NNLO result as an alternative to the NLO+NNLL resummed result. The advantage of the NLO+NNLL resummed results is that they contain powers of $\alpha_s$ to all orders, while the advantage of the approximate NNLO results is that we used the soft function to one order higher (NNLO) and the results do not contain higher orders of $\alpha_s$ which are sometimes not desired. In any case, the two results are extremely close to each other, and we recommend our approximate NNLO result as the most precise theoretical results available to be compared with future experiments, because it turns out to have smaller scale variations.

 In Section~\ref{basics.sec}, we review the formalism and SCET resummation in the threshold region.  This follows closely
 the approach of  Ref. \cite{Ahrens:2010zv}.
 Section~\ref{results.sec} contains results for the NNLL differential and total cross sections, along with approximate results for the NNLO cross 
 section for $pp\rightarrow W^+W^-$. Brief conclusions are presented in Section~\ref{conc.sec}. 
\section{Basics}
\label{basics.sec}
In this section, we review the fixed order results for $pp\rightarrow W^+W^-$ (\ref{born} and \ref{nlo}), the SCET formalism used
to derive the RG improved NNLL results for the differential and total cross sections, including the matching to the fixed order NLO result (\ref{thresh_res}), and the derivation of an approximate NNLO result(\ref{approxnnlo}).

\subsection{Born Level Result}
\label{born}

\begin{figure}[tb]
\centering
\subfigure[]{\includegraphics[height=0.15\textheight,clip]{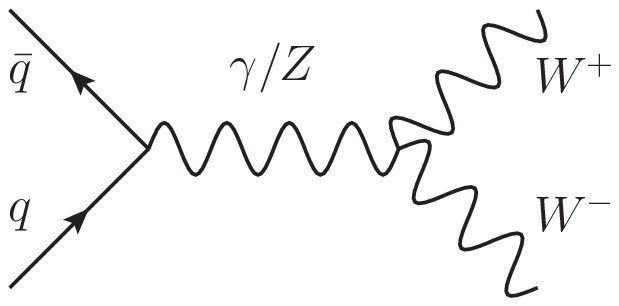}}
\subfigure[]{\includegraphics[height=0.15\textheight,clip]{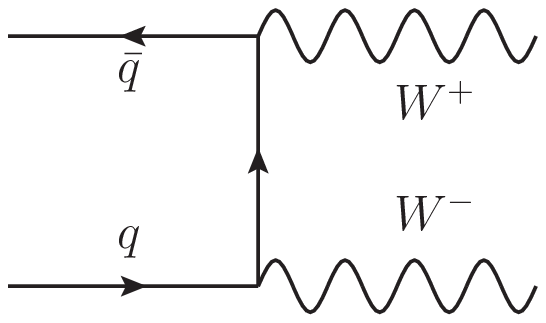}}
\caption{Feynman diagrams for the (a) $s$-channel and (b) $t$-channel contributions to $q\bar{q}\rightarrow W^+W^-$.}
\label{feyn.FIG}
\end{figure}

The Born level process arises through the annihilation process 
\begin{eqnarray}
q(p_1)+{\overline q}(p_2)\rightarrow W^+(p_3)+W^-(p_4)\, .
\end{eqnarray}
  This annihilation proceeds via $s$-channel $\gamma/Z$ exchange and a $t$-channel contribution, as shown in Figs.~\ref{feyn.FIG}(a) and (b), respectively.  The partonic cross section is
\begin{eqnarray}
{\hat \sigma}^0_{q{\overline q}} &=& {1\over 2 s}\int d\Phi_2 \mid A^0_{q{\overline q}}(s,t)\mid^2\, ,
\end{eqnarray}
where the partonic level invariants are
\begin{eqnarray}
s&=& (p_1+p_2)^2\nonumber \\
t&=& (p_1-p_3)^2=M_W^2-{s\over 2}(1-\beta\cos\theta)\nonumber \\
u&=& (p_1-p_4)^2\, ,
\end{eqnarray} 
with $s+t+u=2M_W^2$ and $\beta=\sqrt{1-{4M_W^2\over s}}$.  At the  Born level we have $M_{WW}^2=s$. The two body phase space is
\begin{eqnarray}
d\Phi_2&=&
{1\over 8 \pi s} dt={\beta\over 16\pi } d\cos\theta\, .
\end{eqnarray}
Finally, the color-averaged and spin-summed and averaged matrix element squared is
\begin{equation}
\mid A^0_{q{\overline q}}\mid^2={1\over 4 N_C} 
\biggl\{
c_{q}^{tt}F_q^0(s,t)+c_{q}^{ss}(s)K_q^0(s,t)-c_q^{ts}(s)J_q^0(s,t)\biggr\}
\, ,
\label{lores}
\end{equation}
where $c_{q}^{ss}K_q^0$ is the $s$-channel contribution, $c_{q}^{tt}F_q^0$ is the $t$-channel contribution, and $c_q^{ts}J_q^0$ is from $s$- and $t$-channel interference.  The results have been found in Ref. \cite{Frixione:1993yp} and are given in Appendix ~\ref{appa} for convenience. 

Due to the collinear factorization~\cite{Collins:1989gx,Bodwin:1984hc,Collins:1985ue,Collins:1988ig,Aybat:2008ct}, the hadronic level cross section is obtained by convolving the partonic level cross section with PDFs.  In general, the hadronic cross section can be written as
\begin{eqnarray}
\displaystyle {d^2\sigma\over dM_{WW} d\cos\theta}
=\sum_{ij}
{\beta_W\over 16 \pi M_{WW}  S}
\int _\tau^1 
{dz\over z}
{\cal{ L}}_{ij}
\biggl({\tau\over z}, \mu_f\biggr)
C_{ij}(z,M_{WW},\cos\theta,\mu_f)
\label{had_born}
\end{eqnarray}
where the sum runs over all possible initial state partons, $\mu_f$ is the factorization scale,
\begin{equation}
\tau = \frac{M^2_{WW}}{S},\quad z=\frac{M^2_{WW}}{s},\quad{\rm and}\quad \beta_W=\sqrt{1-\frac{4 M^2_W}{M^2_{WW}}}\,.
\end{equation}
The long-distance collinear physics is described by the parton luminosity
\begin{equation}
{\cal L}_{ij}\biggl(y,\mu_f\biggr)=
\int_y^1{dx\over x}f_i(x,\mu_f)f_j\biggl({y\over x},\mu_f\biggr),
\end{equation}
where $f_i$ is the PDF of a parton with flavor $i$.  The coefficient functions describe the hard partonic process and can be written as a power series in $\alpha_s$,
\begin{equation}
C_{ij}=C_{ij}^0+{\alpha_s\over 4\pi}C_{ij}^1+\ldots
\, .
\label{coeffs}
\end{equation}
We have chosen the normalization of Eq.~(\ref{had_born}) such that the leading order coefficient is
\begin{equation}
C_{ij}^0=\delta(1-z)\mid A_{ij}^0\mid^2\, ,
\end{equation}
and the Born level cross section is recovered.

\subsection{NLO result}
\label{nlo}
At NLO,  the  scattering coefficients of Eq.~(\ref{had_born}) receive corrections from virtual loops and real gluon emission in the
$q {\overline q}$ channel, along with tree level contributions from $qg\rightarrow q W^+W^-$. In dimensional regularization with $N=4-2\epsilon$, the one-loop virtual diagrams contribute
\begin{equation} 
\sigma_{VIRT}={1\over 2 s}\int d\Phi_2\mid A^1_{q{\overline q}}(s,t)\mid^2\, ,
\end{equation}
where
\begin{eqnarray}
\mid A_{q{\overline q}}^1\mid^2&=&
\mid A_{q{\overline q},reg}^1\mid^2+\mid A_{q{\overline q},div}^1\mid^2
\\
\mid A_{q{\overline q},div}^1\mid^2 &=&-
{\alpha_s\over 4\pi}C_F 
\biggl({4\pi\mu^2\over s}\biggr)^\epsilon
\Gamma(1+\epsilon)
\biggl({4\over \epsilon^2}+{6\over \epsilon}\biggr)\mid
A^0_{q{\overline q}}(s,t)\mid^2\nonumber \\
\mid A_{q{\overline q},reg}^1\mid^2&=&{\alpha_s\over 16 \pi N_C}C_F
\biggl\{
c_{q}^{tt}F_q^1(s,t)+c_{q}^{ss}(s)K_q^1(s,t)-c_q^{ts}(s)J_q^1(s,t)\biggr\}
\, .
\label{nlores}
\end{eqnarray}
We note that since QCD does not renormalize electroweak couplings, all the UV divergences cancel  leaving only IR divergences in $A^1_{q\bar{q},div}$.  The one-loop corrections to the $t$-channel exchange are given by $c_{q}^{tt}F_q^1(s,t)$, to the $s$-channel exchange by $c_{q}^{s s}K_q^1(s,t)$, and the interference  between the $s$- and $t$-channels by $c_q^{ts}J_q^1(s,t)$. Expressions for these terms can be found in  Appendix~\ref{appa}.  As will be discussed in the next section, the real hard gluon emission contribution is not relevant for the resummation of the threshold logarithms and we therefore do not give it here, although it can be found in Ref. \cite{Frixione:1993yp}.

\subsection{Threshold Resummation and Matching}
\label{thresh_res}

We now discuss the resummation of the large logarithms in the partonic threshold limit, $z\rightarrow 1$.  In this limit, since there is no phase space available for hard gluon emission, the total phase space is well described by the Born $2\rightarrow 2$ process and Eq.~(\ref{had_born}) can be used.  In addition to the collinear factorization, in the threshold limit the coefficient functions can be further factorized into hard, $H$, and soft, $S$, functions:\footnote{Since we are interested in a color singlet final state, the soft function $S$ has no $\cos\theta$ dependence.}
\begin{equation}
C_{q{\overline q}}(z,M_{WW},\cos\theta,\mu_f)\equiv H(M_{WW},\cos\theta,\mu_f)S(\sqrt{s}(1-z),\mu_f)+
{\cal O}(1-z)\, .
\label{hsdef}
\end{equation}
The soft function is given by the vacuum expectation values of soft Wilson loops~\cite{Becher:2006mr,Becher:2007ty} and the hard function is calculated by matching the  full QCD result onto the relevant SCET operator.  It is this matching that integrates out the hard QCD modes and leaves the soft and collinear modes that comprise SCET.

The  hard function is given by\footnote{In Eqs. (\ref{wilscoeff}) and (\ref{zren}), the sum over Dirac structures is implied.  See Ref. \cite{Ahrens:2011mw} for
an example of the relevant notation.}
\begin{eqnarray}
H(M_{WW},\cos\theta,\mu)=|C_{WW}(M_{WW},\cos\theta,\mu)\mathcal{O}_{WW}|^2\, ,
\label{wilscoeff}
\end{eqnarray}
where $C_{WW}$ is the Wilson coefficient of the relevant SCET operator $\mathcal{O}_{WW}$.  The Wilson coefficient is calculated by matching the renormalized QCD and SCET amplitudes:
\begin{eqnarray}
\mathcal{M}^{\rm ren}(\epsilon,M_{WW},\cos\theta)=Z(\epsilon,M_{WW},\mu)\, C_{WW}(M_{WW},\cos\theta,\mu)\mathcal{O}_{WW},
\label{zren}
\end{eqnarray}
where $\mathcal{M}^{\rm ren}$ is the renormalized QCD amplitude and $Z$ is the SCET renormalization constant. In dimensional regularization, SCET loops are scaleless and vanish.  This implies that the UV and IR singularities of SCET coincide and cancel.  Since SCET and QCD describe the same low-scale physics and have the same IR pole structure, $Z$ can be determined by the behavior of IR singularities in QCD~\cite{Giele:1991vf,Kunszt:1994np,Catani:1996jh,Catani:1996vz,Catani:1998bh,Becher:2009cu,Becher:2009qa}.  In the $\overline{\rm MS}$ scheme, we have
\begin{eqnarray}
Z(\epsilon,M_{WW},\cos\theta,\mu) = 1-\frac{\alpha_s C_F}{2\pi}\left(4\pi\right)^\epsilon e^{-\epsilon\gamma_E}\left\{\frac{1}{\epsilon^2}+\frac{1}{\epsilon}\left(\ln\frac{\mu^2}{-M^2_{WW}}+\frac{3}{2}\right)\right\}\, .
\label{SCETren}
\end{eqnarray}
The poles in $Z$ and the NLO QCD squared amplitudes in Eq.~(\ref{nlores}) cancel.  Hence, the one-loop contribution to the hard function is just the finite terms of Eq.~(\ref{nlores})~\cite{Manohar:2003vb,Becher:2006nr}.

Since the hard function is calculated in the perturbative region of QCD it can be expanded in powers of $\alpha_s$:
\begin{equation}
H(M_{WW},\cos\theta,\mu_h)=
H^0(M_{WW},\cos\theta)+{\alpha_s(\mu_h)\over 4\pi}
\biggl[H^1_{reg}(M_{WW},\cos\theta,\mu_h)+
H^1_{extra}(M_{WW},\cos\theta,\mu_h)\biggr],
\label{HNLO}
\end{equation}
where $\mu_h$, termed the hard scale, is the scale at which QCD and SCET are matched.    The normalization of the hard function is such that
\begin{equation}
H^0(M_{WW},\cos\theta)=\mid A^0_{q{\overline q}}\mid^2\, ,
\label{H0}
\end{equation}
and
\begin{eqnarray}
{\alpha_s\over 4\pi} H^1_{reg}(M_{WW},\cos\theta,\mu_h)&=&\mid A_{q{\overline q},reg}^1\mid^2
\nonumber \\
H^1_{extra}(M_{WW},\cos\theta,\mu_h)&=&-
C_F H^0(M_{WW},\cos\theta)\nonumber\\
&&\times\biggl\{{\pi^2\over 3}+
2\log^2\biggl({M_{WW}^2\over \mu_h^2}\biggr)
-6\log\biggl({M_{WW}^2\over \mu_h^2}\biggr)\biggr\}\, .
\label{H1}
\end{eqnarray}

Now we have all the pieces to resum the threshold logarithms.  As mentioned before, the hard function is calculated at the matching scale $\mu_h$.  Since the soft function describes the soft physics, it is evaluated at a soft scale, $\mu_s$, associated with the scale of soft gluon emission.  By using the RG equations (RGEs), the hard and soft functions can be evolved to the factorization scale.  The RG evolution of the soft function resums logs of the form $\alpha_s^n\ln^m(\mu_s/M_{WW})$.  By choosing the soft scale $\mu_s\sim M_{WW}(1-\tau)$, the RGE running resums the large threshold logarithms.  In Table~\ref{scetorders} we list the accuracy of the resummation at a given order. The resulting coefficient function is~\cite{Becher:2007ty}
\begin{eqnarray}
C(z,M_{WW},\cos\theta,\mu_f)&=&
U(M_{WW},\mu_h,\mu_s,\mu_f)\,H(M_{WW},\cos\theta,\mu_h)
\nonumber\\
&&\times{\tilde s}\biggl(
\ln\biggl({M_{WW}^2\over \mu_s^2}\biggr)+\partial_\eta,\mu_s\biggr) \,
{e^{-2\gamma_E\eta}\over\Gamma(2\eta)}
{z^{-\eta}\over (1-z)^{1-2\eta}}
\end{eqnarray}
where $\eta=2a_\Gamma(\mu_s,\mu_f)$ and
\begin{eqnarray}
U(M_{WW},\mu_h,\mu_s,\mu_f) = \left(\frac{M^2_{WW}}{\mu^2_h}\right)^{-2a_\Gamma(\mu_h,\mu_s)}\exp\left[4S(\mu_h,\mu_s)-2a_{\gamma^V}(\mu_h,\mu_s)+4a_{\gamma^\phi}(\mu_s,\mu_f)\right].
\end{eqnarray}  
  Finally, ${\tilde s}$ is can be expressed as a power expansion in logarithms,
 \begin{eqnarray}
{\tilde s}(L,\mu)&=&1+
{\alpha_s\over 4 \pi}\sum_{n=0}^2s^{(1,n)}L^n
+\biggl({\alpha_s\over 4 \pi}\biggr)^2
\sum_{n=0}^4 s^{(2,n)}L^n.
\end{eqnarray}
Expressions for the $s^{(i,n)}$ are given in Appendix~\ref{appb} and are identical to those found for Drell-Yan~\cite{Becher:2007ty}.

We will present results both at NLL and NNLL.  The corresponding order of the needed functions is given in Table \ref{scetorders}.  
Explicit expressions for the functions $a$ and $S$, and the anomalous dimensions $\Gamma$, $\gamma^V$, and $\gamma^\phi$ can be found in the appendices of Ref. \cite{Becher:2007ty}. The choice of the soft scale, $\mu_s$ ,is discussed in Section \ref{difs}.

The resummed results are only valid in the region $z\rightarrow 1$.  To extend these results to all $z$, the resummed cross section needs to be matched with the full fixed order cross section. This allows the inclusion of the non-singular terms in $(1-z)$  which are present in the fixed order result but not the resummed result.  For NNLL resummation, this means matching with the NLO cross section:
\begin{eqnarray}
d\sigma^{NLO+NNLL}&\equiv & d\sigma^{NNLL}(\mu_h,\mu_f,\mu_s)+\biggl(d\sigma^{NLO}(\mu_f)-d\sigma^{NLO}(\mu_f)\mid_{leading~singularity}\biggr)
\label{matchdef}
\end{eqnarray}
where $d\sigma^{NNLL}$ is the threshold resummed cross section and $d\sigma^{NLO}\mid_{leading~singularity}$ contains only the ${\cal O}(\alpha_s)$ NLO terms which are singular as $z\rightarrow 1$,
\begin{equation}
d\sigma^{NLO}\mid_{leading~singularity}\equiv d\sigma^{NNLL}\mid_{\mu_h=\mu_f=\mu_s}\, .
\label{matching}
\end{equation}
Subtracting $d\sigma^{NLO}\mid_{leading~singularity}$ prevents double counting of terms common to the resummed and fixed order results.  Also, in the limit $z\rightarrow 1$, the matched cross section corresponds to the resummed results, as desired.

\begin{table}[tp]
\renewcommand{\arraystretch}{1.4}
\centering
\caption{Accuracy of the SCET resummation at a given order and required accuracy of SCET inputs.}
\begin{tabular}{|c|c|c|c|c|}
\hline
Order & Accuracy: $\alpha_s^n \ln^m(\mu_s/M_{WW})$& $\Gamma_{cusp}$ &$\gamma^h,\gamma^\phi$ & $H, {\tilde s}$ \\
\hline
NLL & $2n-1 \leq m \leq 2n$& 2-loop& 1-loop& tree\\
NNLL& $2n-3 \leq m \leq 2n$&3-loop&2-loop&1-loop \\
\hline
\end{tabular}
\vspace{-1ex}
\label{scetorders}
\end{table}
\subsection{Approximate NNLO Results}
\label{approxnnlo}
The full NNLO cross section can only be determined from a complete calculation.  However, the scale dependent terms that are singular as $z\rightarrow1$ can be determined to NNLO accuracy via the known hard and soft functions and their respective RGEs.  As will be shown in the next section, most of the NLO correction comes from the leading singular piece.  Hence, we expect that including the scale dependent, leading singular pieces of the NNLO cross section is a good estimate of the full NNLO result.  The inclusion of these pieces is known as approximate NNLO.

The coefficient function in Eq.~(\ref{hsdef}) can be expanded in a power series,
\begin{eqnarray}
C(z,M,\cos\theta,\mu)&=&C^{0}(z,M,\cos\theta,\mu)+\frac{\alpha_s}{4\pi}C^{1}(z,M,\cos\theta,\mu)
\nonumber \\ &&
+\left(\frac{\alpha_s}{4\pi}\right)^2C^{2}(z,M,\cos\theta,\mu)
\, .
\end{eqnarray}
The leading order, $C^{0}$, and NLO, $C^{1}$, contributions are fully known analytically.  The NNLO contribution, $C^{2}$, can only be approximately determined from the hard and soft functions.  The soft function is known fully to NNLO~\cite{Becher:2007ty}; hence the only approximation comes from the unknown scale-independent NNLO piece of the hard function.  The approximate NNLO cross section is found by calculating the scale-dependent, leading singular pieces of $C^{2}$ and adding this contribution to the full NLO result.

We expand the hard function as a power series in $\alpha_s$:
\begin{eqnarray}
H_{approx}(M_{WW},\cos\theta,\mu_f)=H(M_{WW},\cos\theta,\mu_f)+\left(\frac{\alpha_s}{4\pi}\right)^2H^{2}_{approx}(M_{WW},\cos\theta,\mu_f),
\label{Happrox}
\end{eqnarray}
where the full NLO hard function, $H$, is given in Eq.~(\ref{HNLO}).  The NNLO hard piece, $H^{2}_{approx}$, contains only the scale dependent pieces at NNLO.  We further expand $H^{2}_{approx}$ in a power series of logs:
\begin{eqnarray}
H^{2}_{approx}(M_{WW},\cos\theta,\mu_f)=\sum^3_{n=1}h^{(2,n)}L_{WW}^n,
\label{H2}
\end{eqnarray}
where $L_{WW}=\ln(M^2_{WW}/\mu^2_f)$. The coefficients $h^{(2,n)}$ can be found in Appendix~\ref{appb}.

The approximate NNLO hard function in Eq.~(\ref{Happrox}) is independent of scale to order $\alpha_s^3$.  Hence, scale variation only contains the $\mathcal{O}(\alpha_s^3)$ uncertainties, not taking into account the unknown NNLO scale independent and nonsingular in $(1-z)$ pieces at $\mathcal{O}(\alpha_s^2)$.  Variation of the factorization scale may therefore underestimate the total uncertainty in the approximate NNLO result. However, this uncertainty can be further estimated by noting that there is an ambiguity in the logs used to expand $H^{2}$.  For example, introduce a scale $Q_h\sim M_{WW}$.  Then $H^{2}$ can be expanded in logs of the form $L_Q=\ln(Q^2_h/\mu^2_f)$ instead of $L_{WW}$.  The difference between these two schemes will be order one contributions to the scale independent piece.  Hence, in addition to the variation of the factorization scale, the uncertainty associated with the unknown NNLO scale independent piece is estimated by making the replacement $L_{WW}\rightarrow L_Q$ in Eq.~(\ref{H2}) and varying the new scale $Q_h$ around the central value $M_{WW}$, which is the natural choice from the RGEs.  

We note that when including the scale independent pieces, the full NNLO hard function is independent of the scale $Q_h$.  Similarly, since the soft function is known fully to NNLO there is no ambiguity in the choice of scales used in the log expansion.

\section{Results}
\label{results.sec}
\subsection{Soft Scale Choice}
\label{scale}
In the process of performing resummation in the SCET formalism, two additional scales are introduced: the hard scale $\mu_h$, where the hard function is evaluated; and the soft scale, $\mu_s$, where the soft function is evaluated.  Since the hard function is calculated from matching QCD onto SCET at the scale of the hard scattering process, the central value of $\mu_h$ is naturally chosen to be the scale of the hard scattering process, $\mu_h=M_{WW}$.  

The soft scale is chosen to be associated with the scale of the soft gluon emissions, such that the RG evolution resums large logs associated with soft gluon emission at threshold. Following Refs.~\cite{Becher:2006nr,Becher:2006mr,Becher:2007ty}, we choose the soft scale to be related to the hadronic energy scale, avoiding the Landau poles that plague the traditional perturbative QCD resummation~\cite{Catani:1996yz,Becher:2006nr}.  Hence, in the hadronic threshold limit $\tau\rightarrow1$ we want $\mu_s\sim M_{WW}(1-\tau)$.  However, at hadron colliders most of the cross section is accumulated far from $\tau=1$ and the choice of soft scale away from this limit is less clear.

\begin{figure}[tb]
\subfigure[]{
      \includegraphics[width=.45\textwidth,angle=0,clip]{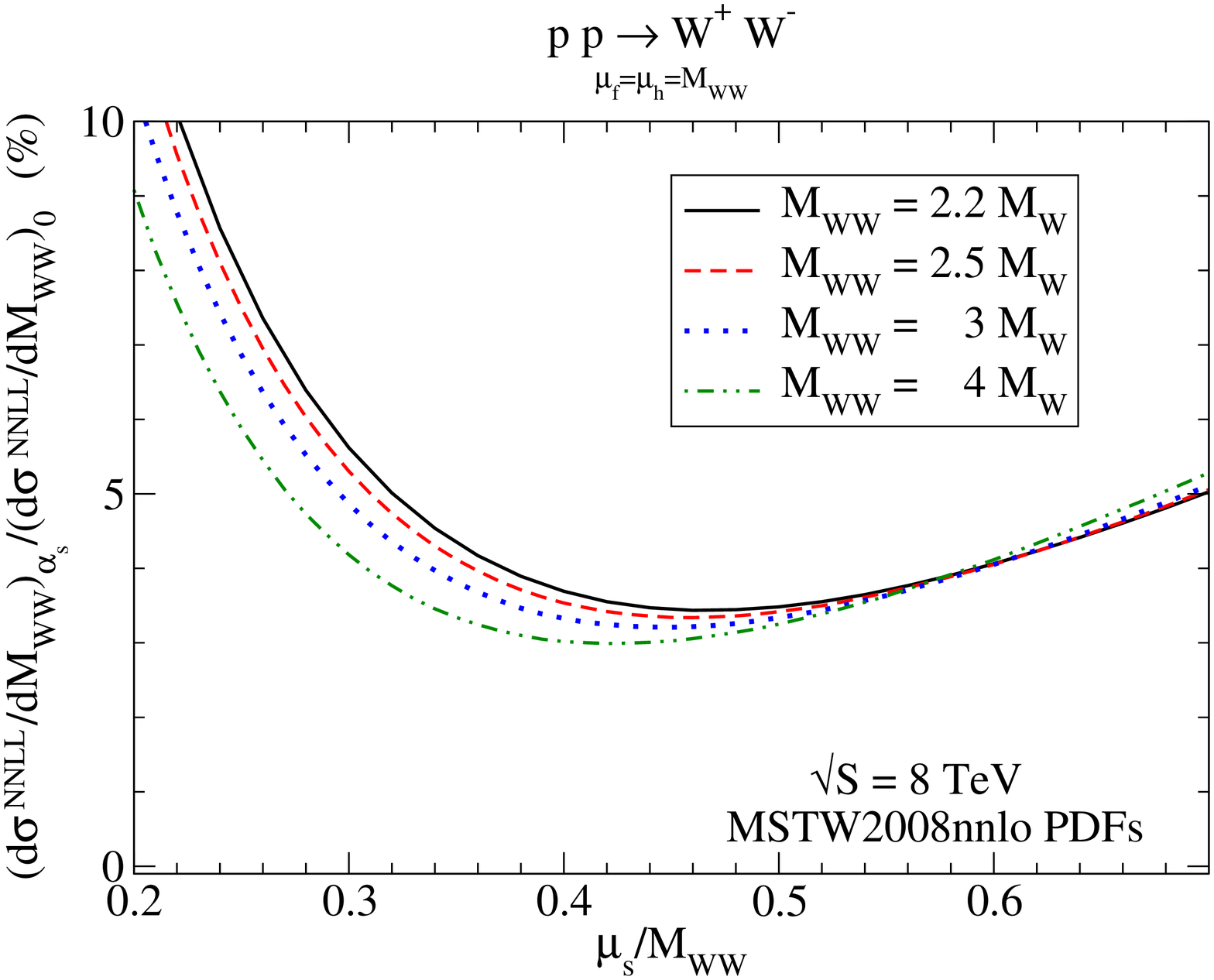}
      \label{musdep.FIG}}
\subfigure[]{
      \includegraphics[width=0.45\textwidth,clip]{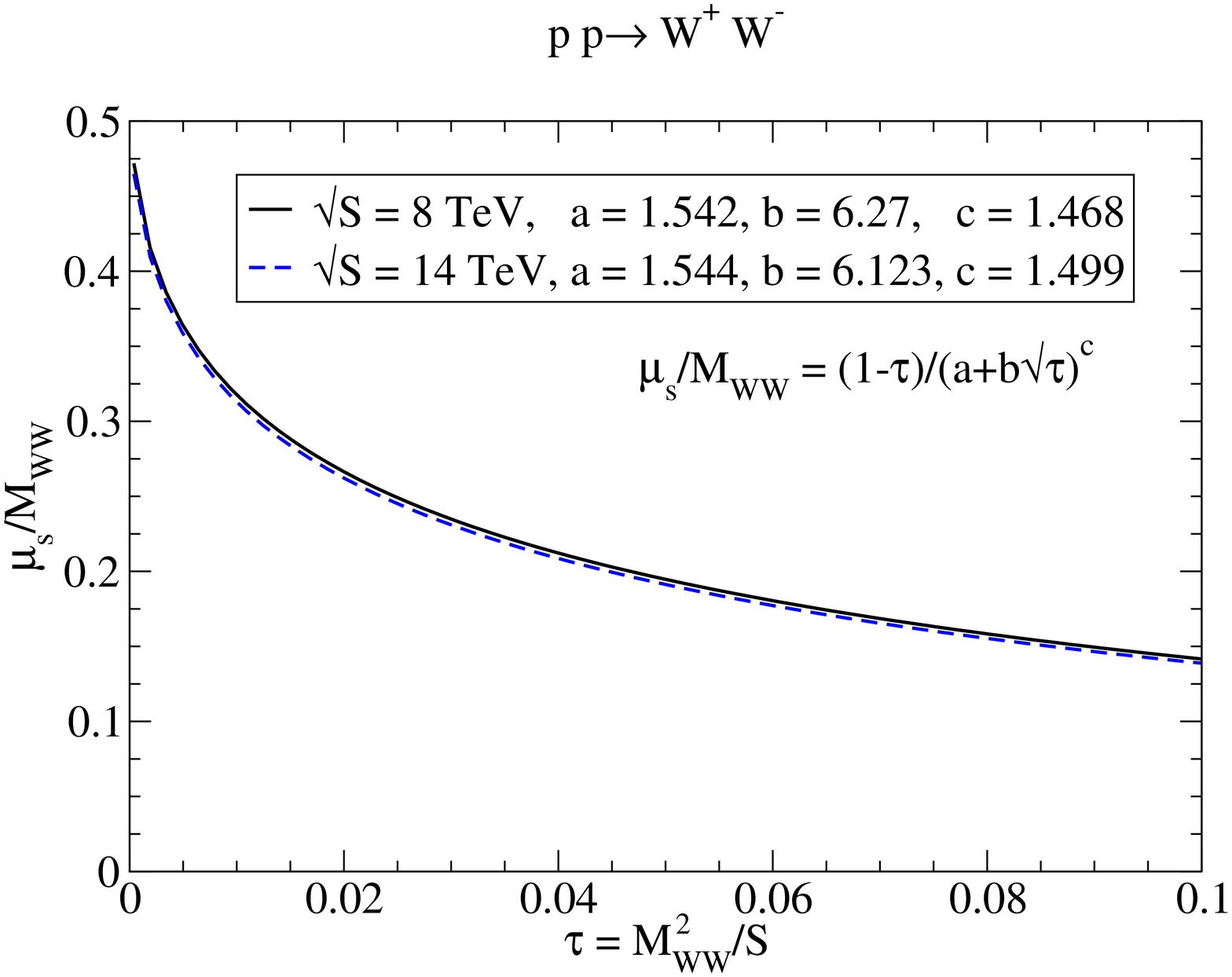}
      \label{musChoice.FIG}}
\caption{(a) The ratio of the NNLL-resummed invariant mass distribution evaluated with only the $\mathcal{O}(\alpha_s)$ piece of the soft function, $(d\sigma^{NNLL}/dM_{WW})_{\alpha_s}$, to the NNLL-resummed distribution evaluated with only the LO piece of the soft function, $(d\sigma^{NNLL}/dM_{WW})_0$, at $\sqrt{S}=8$~TeV and various values of $M_{WW}$. (b) The minimum value of $\mu_s/M_{WW}$ at $\sqrt{S}=8$~TeV (solid) and $\sqrt{S}=14$~TeV (dashed) as a function of $\tau$.}
\label{mus.FIG}
\end{figure}

Another constraint on $\mu_s$ is that the soft function should be a well-behaved perturbative series.  Hence, away from the threshold region, we choose $\mu_s$ such that the $\mathcal{O}(\alpha_s)$ piece of the soft function is minimized relative to the LO piece.  Figure~\ref{musdep.FIG} shows the ratio of the NNLL-resummed invariant mass distribution evaluated with only the $\mathcal{O}(\alpha_s)$ piece of the soft function, $(d\sigma^{NNLL}_{WW})_{\alpha_s}$, to the NNLL-resummed distribution evaluated with only the LO piece of the soft function, $(d\sigma^{NNLL}_{WW})_0$.  In both distributions, all other pieces of the resummed cross sections (aside from the soft function) are evaluated to full NNLL order.  This ratio is shown for $\sqrt{S}=8$~TeV at various values of $M_{WW}$ with $\mu_f=\mu_h=M_{WW}$ and MSTW2008nnlo PDFs.  The soft scale is chosen to correspond to the minimum of these ratios, which is found to be well described by the parameterization
\begin{eqnarray}
\mu_s = M_{WW} \frac{(1-\tau)}{(a+b\sqrt{\tau})^c}.
\end{eqnarray}
For $\sqrt{S}=8$~TeV, it is found that $a=1.542$, $b=6.27$, and $c=1.468$.  Performing a similar fit at $\sqrt{S}=14$~TeV it is found that $a=1.544$, $b=6.123$, and $c=1.499$.  With this parameterization $\mu_s$ has the correct dependence on $\tau$ in the threshold region.

Figure~\ref{musChoice.FIG} shows the minimum value of the ratio $\mu_s/M_{WW}$ at $\sqrt{S}=8$~TeV (solid) and $\sqrt{S}=14$~TeV (dashed) as a function of $\tau$.  As can be clearly seen, the hadronic energy makes little difference to the choice of the soft scale.  For simplicity, independent of the hadronic energy scale, all results presented here use the central value of the soft scale corresponding to the 8~TeV solution:
\begin{eqnarray}
\mu_s^{\rm min} = M_{WW} \frac{(1-\tau)}{(1.542+6.27\sqrt{\tau})^{1.468}}.
\end{eqnarray}

\subsection{Differential Cross Section}
\label{difs}

\begin{figure}[tb]
      \includegraphics[width=.6\textwidth,angle=0,clip]{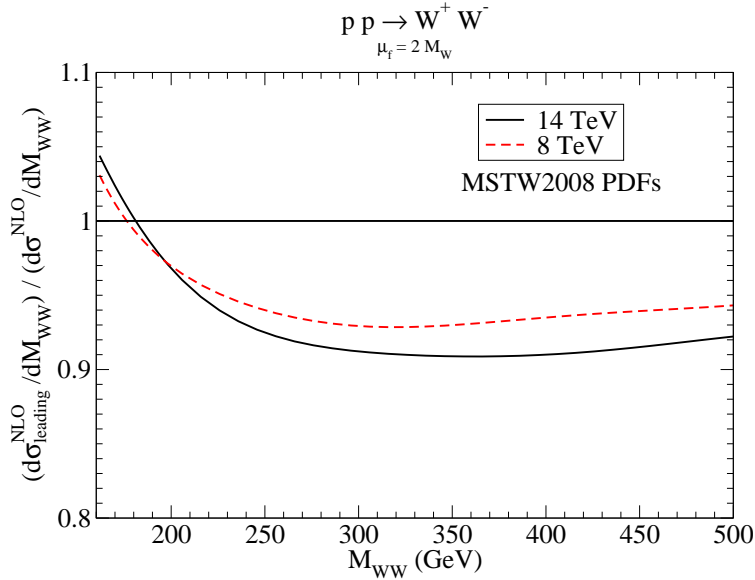}
\caption{Ratio of the contribution of the leading singularity to the fixed order NLO cross section at $\sqrt{S}=8$ and $14$ TeV using MSTW2008 PDFs.
NNLO (NLO) PDFs are used for the NNLL leading contribution (fixed order NLO contribution). }
\label{leadingfig}
\end{figure}

We begin by considering the validity of the matching of the  NNLL results to the fixed order NLO results. In order for the matching of Eq.~(\ref{matchdef}) to be valid, the sub-leading terms in $(1-z)$ must be small. In Fig. \ref{leadingfig}, we show the contribution of the leading singularity to the fixed order NLO differential cross section for $\sqrt{S}=8$~TeV and $\sqrt{S}=14$~TeV. We fix  the central scale to be $\mu_f=2M_W$. From this figure, we see that the leading singularity captures $\sim 90\%$ of the NLO fixed order cross section.  Hence, the threshold singularities contribute most of the NLO cross section and we may expect that by resumming the higher order logarithms we capture most of the higher order cross section.

\begin{figure}[tb]
\subfigure[]{
      \includegraphics[width=0.45\textwidth,angle=0,clip]{MWW_14TeV.eps}\label{mww_14_all}}
\subfigure[]{
      \includegraphics[width=0.45\textwidth,clip]{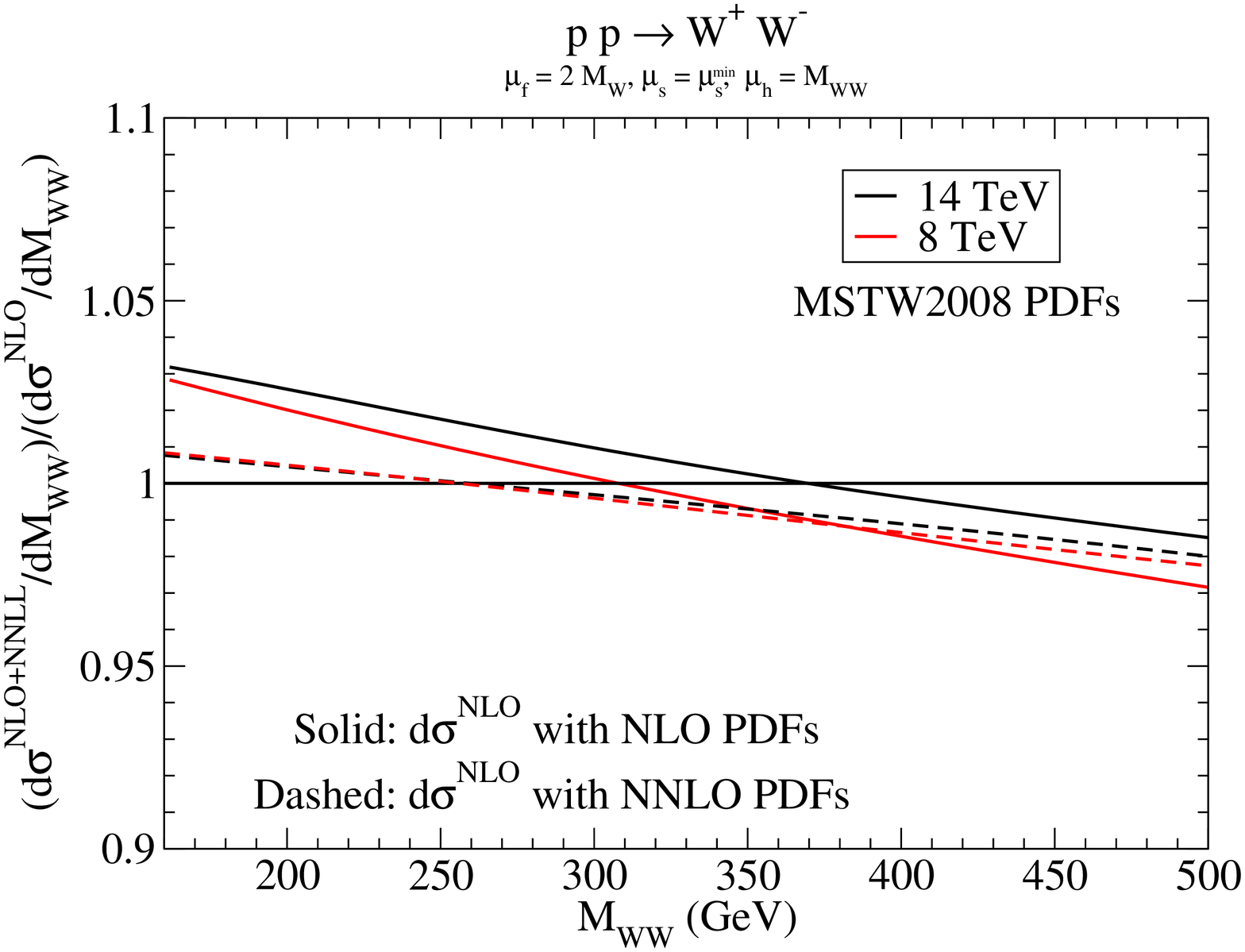}\label{NLO_ratio}}
\caption{(a) Invariant mass distribution and factorization scale dependence of fixed order and matched differential cross sections at $\sqrt{S}=14$~TeV using MSTW2008 PDFs. The dash-dot-dot curves have $\mu_f=M_W$ and the dashed curves have $\mu_f=4M_W$. (b) Ratio of the NLO+NNLL matched and NLO invariant mass distribution for $\sqrt{S}=8$ and $14$ TeV with the NLO cross section with NLO PDFs (solid) and NNLO PDFs (dashed).  The factorization scale is fixed to $\mu_f=2 M_W$.}

\end{figure}

In Fig. \ref{mww_14_all}, we show $d \sigma/dM_{WW}$ versus $M_{WW}$ for $\sqrt{S}= 14$~TeV, with MSTW2008 PDFs.  The curves are LO, NLO, NLL (matched), and NNLL (matched), with $\mu_h=M_{WW}$, $\mu_s=\mu_s^{\rm min}$ and $\mu_f$ varied up and down by a factor of $2$ from the central value of $\mu_f^0=2M_W$.    It is apparent that the NNLL resummation slightly increases the rate at the peak.  

A change in the $W^+W^-$ invariant mass distribution may be consequential to the analysis of the $H\rightarrow W^+W^-\rightarrow 2\ell 2\nu$ decay channel.  In the zero jet bin, the major background is the SM (non-Higgs) production of $W^+W^-$~\cite{CMS:bxa,ATLAS:2013wla}.  To estimate this background a sideband analysis is performed.  A control region is defined with a minimum dilepton invariant mass, where the $W^+W^-$ background strongly dominates the Higgs signal.  The control region is used to normalize the cross section and then Monte Carlo is used to extrapolate the line shapes into the signal region.  If higher order corrections alter the $M_{WW}$ distribution, the dilepton invariant mass distribution will be changed and the extrapolation to the signal region will need to take this into account.

In Fig.~\ref{NLO_ratio} we explore the effect of higher order corrections on the invariant mass distribution by plotting the ratio of the NLO+NNLL matched and NLO $M_{WW}$ distributions.  The scales are set to be $\mu_h=M_{WW}$, $\mu_s=\mu_s^{\rm min}$, and $\mu_f = 2 M_W$.  For the NLO cross section evaluated with NLO PDFs (solid), the resummation increases the invariant mass distribution by $\sim 3-4\%$ in the peak region for both $\sqrt{S}=8$ and $14$ TeV and decreases it by $\sim 2\%$ and $\sim 1\%$ in the high mass region for $\sqrt{S}=8$ and $14$ TeV, respectively.  However, most of this change in the $M_{WW}$ distribution is from the different PDFs used for the resummed and NLO results, as can be seen when the NLO cross section is evaluated with NNLO PDFs (dashed).  In this case, the resummation only alters the invariant mass distribution by $\lesssim 1\%$ for a wider range $M_{WW}$.  This indicates that the calculation of the $W^+W^-$ cross section is firmly under theoretical control.

\begin{figure}[tb]
\subfigure[]{
      \includegraphics[width=.45\textwidth,angle=0,clip]{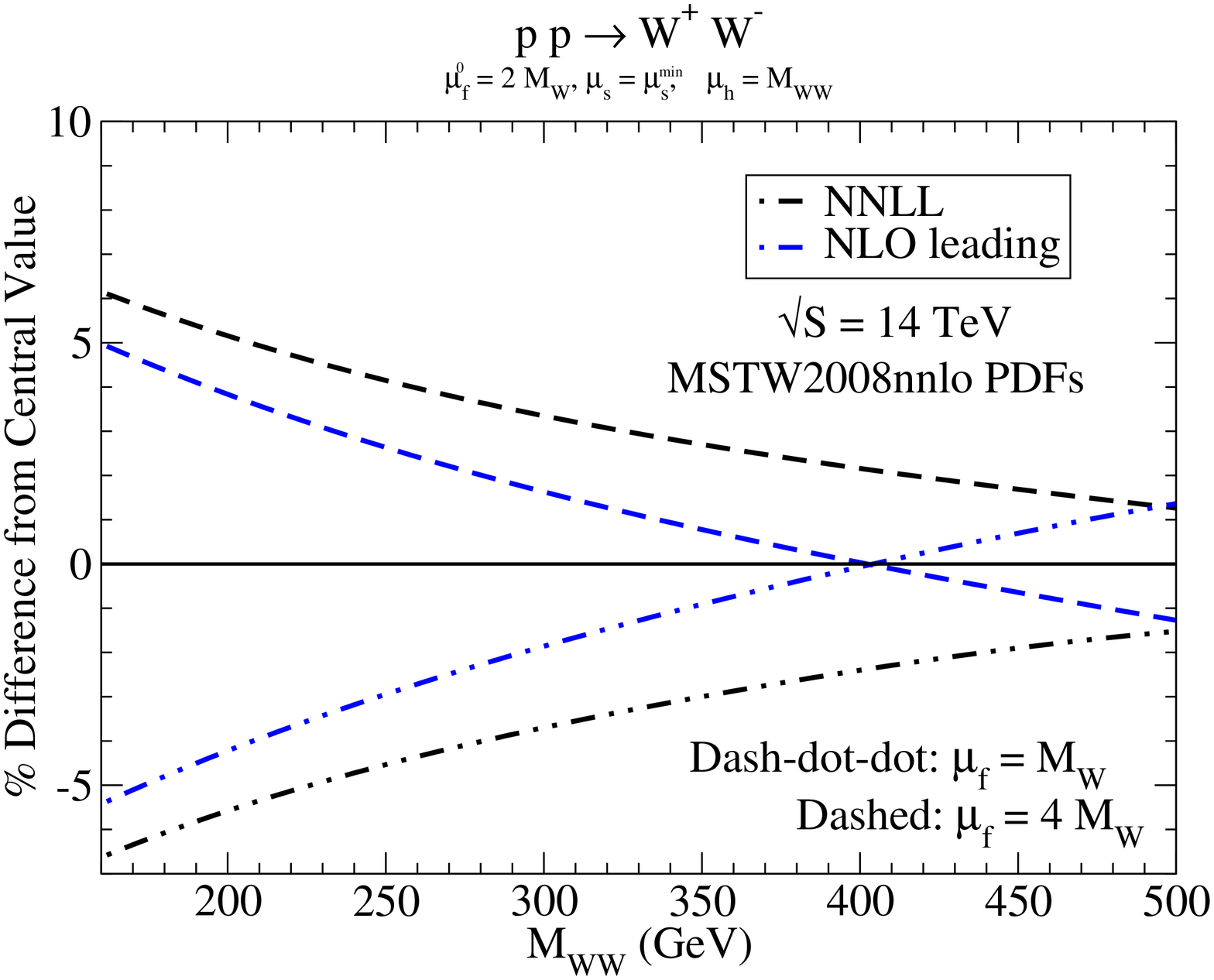}\label{NNLL_lead}}
\subfigure[]{
      \includegraphics[width=.45\textwidth,angle=0,clip]{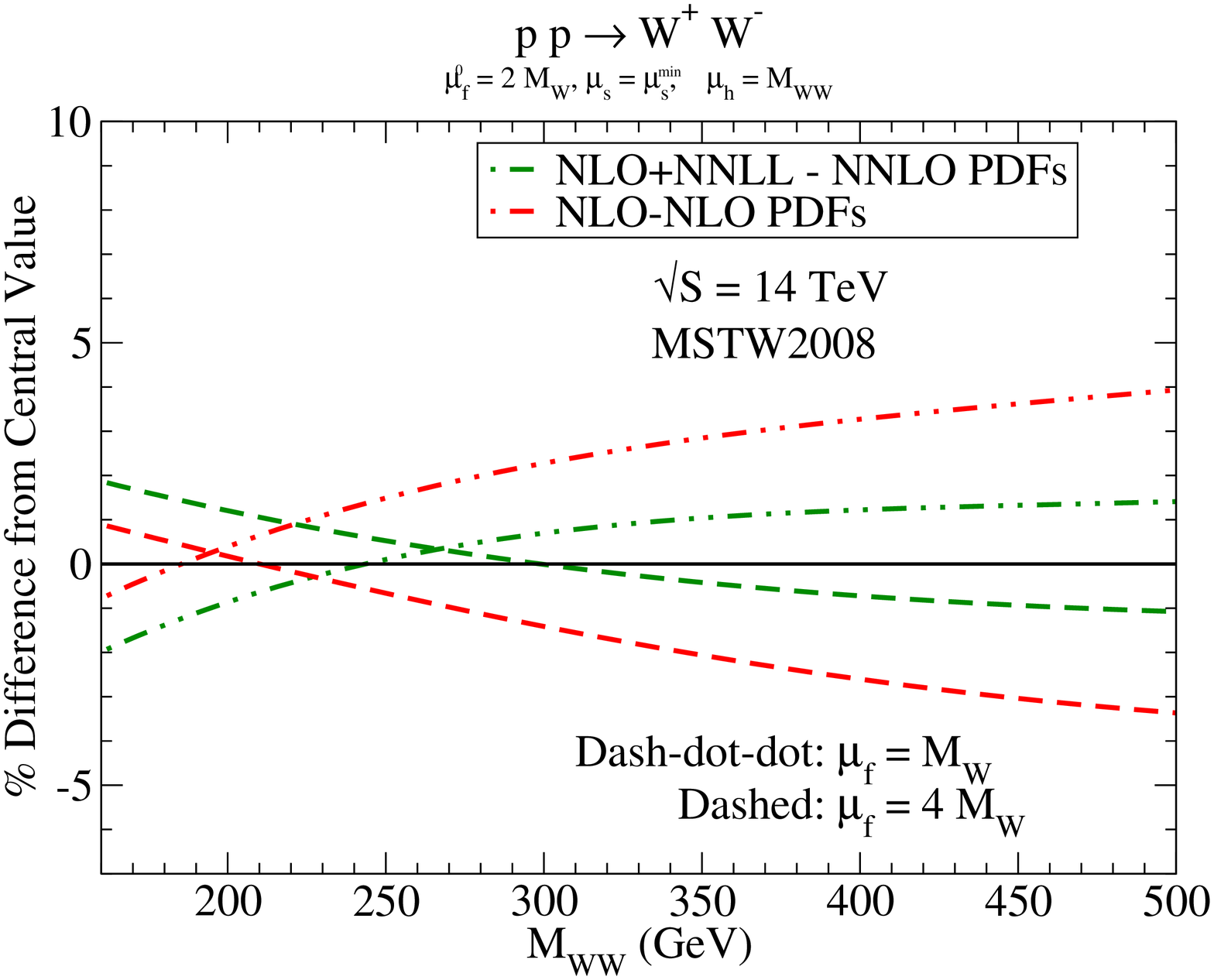}\label{NNLL_NLO}}
\caption{Factorization scale dependence of (a) NNLL resummed and NLO leading singularity, and (b) NLO and matched NNLL invariant mass distributions at $\sqrt{S}=14$~TeV using MSTW2008 PDFs. The dash-dot-dot curves have $\mu_f=M_W$ and the dashed curves have $\mu_f=4M_W$.
}
\label{matched_nnll}
\end{figure}

The factorization scale dependence of the invariant mass distributions is shown in Fig. \ref{matched_nnll} for (a) the NNLL resummed and NLO leading, and (b) the NLO and NNLL matched results.  Here we present the factorization scale dependence as a percent difference from the central value. Using the definition of matching in Eq.~(\ref{matchdef}), Fig.~\ref{NNLL_lead} indicates that there is a cancellation between the $\mu_f$ dependencies the NNLL resummed and NLO leading results for $M_{WW}\lesssim 400$~GeV.  Comparing Figs. \ref{matched_nnll} (a) and (b), it is apparent that for $M_{WW}\gtrsim190$~GeV the NLO $\mu_f$ dependence also cancels against the NNLL resummed dependence.  Hence, although the cancellation is not as efficient at lower $M_{WW}$, as the invariant mass increases the $\mu_f$ dependence of the NNLL matched result is less than that of the NLO result.  This can be seen in Fig.~\ref{NNLL_NLO}, where we see that that for $M_{WW}\gtrsim 220-230$~GeV the $\mu_f$ dependence of the NNLL matched result is lower than the NLO result.  Hence, although the scale dependence of the NNLL matched result is larger than that of the NLO result at the peak of the invariant mass distribution, one can show that the resummation and matching procedure decreases the factorization scale dependence of the total cross section relative to the NLO result.

\begin{figure}[tb]
      \includegraphics[width=.6\textwidth,angle=0,clip]{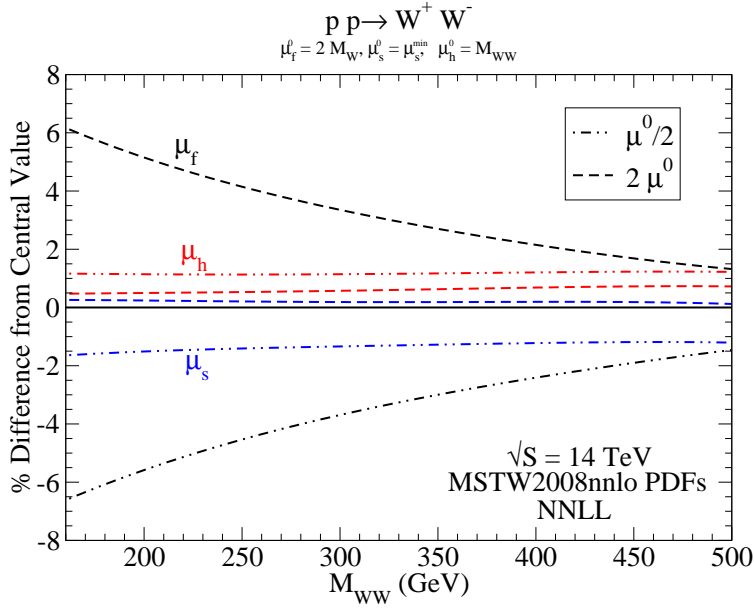}
\caption{Scale dependence of the $\sigma^{NNLL}$ differential cross sections at $\sqrt{S}=14$~TeV using MSTW2008nnlo PDFs.
The scales are varied by a factor of 2 up and down from the central scales, $\mu_h^0=M_{WW}$, $\mu_s^0=\mu_s^{\rm min}$ and $\mu_f^0=2M_W$.}
\label{mww_14_scales}
\end{figure}

In Fig. \ref{mww_14_scales}, we show  the deviation from the central scales for the NNLL resummed differential cross section, $d \sigma^{NNLL}/dM_{WW}$, versus $M_{WW}$ for $\sqrt{S}= 14$~TeV, with MSTW2008nnlo PDFs.   Again, we present the scale dependence as a percent difference from the central value.  The central scales are $\mu_h^0=M_{WW}$, $\mu_s^0=\mu_s^{\rm min}$ and $\mu_f^0=2M_W$  and are  separately  varied up and down by a factor of $2$. The hard and soft scale variations are of the order of $\sim 1-2\%$ and are relatively independent of $M_{WW}$. The factorization scale dependence near the peak is $\sim \pm 6\%$ and is always greater than the hard and soft scale dependencies for the invariant mass range presented.

\begin{figure}[tb]
      \includegraphics[width=.6\textwidth,angle=0,clip]{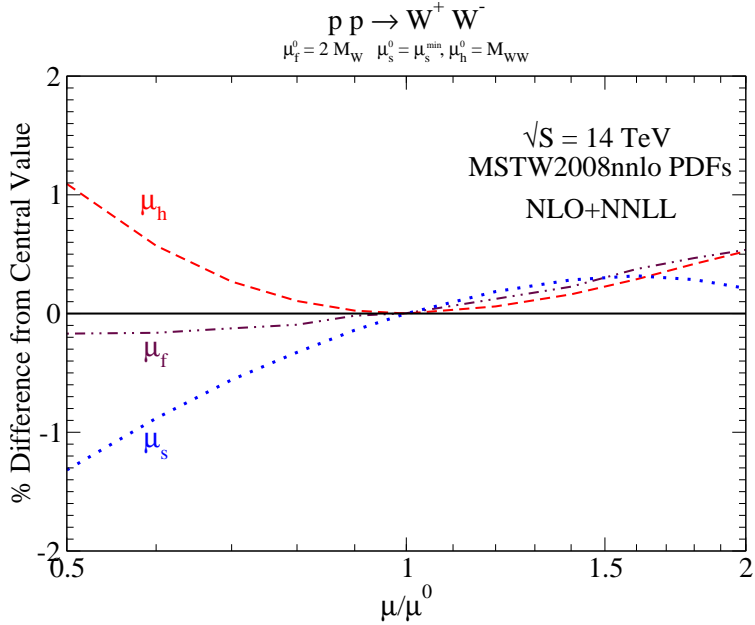}
\caption{Scale dependence of  the matched $\sigma^{NLO+NNLL}$ cross sections at $\sqrt{S}=14$~TeV using MSTW2008nnlo PDFs. The scales are varied by a factor of 2 up and down from the central scales, $\mu_h^0=M_{WW}$, $\mu_s^0=\mu_s^{\rm min}$ and $\mu_f^0=2M_W$.
}
\label{tot_14_scales}
\end{figure}

Finally, in Fig.~\ref{tot_14_scales} we show the variation of the NNLL matched cross section with the central scale choices $\mu_h^0=M_{WW}$, $\mu_s^0=\mu_s^{\rm min}$ and $\mu_f^0=2M_W$ at $\sqrt{S}=14$~TeV. The cross section varies by less than $\pm 2\%$ as the scales are varied from the central values.  Again, we see that there is a large cancellation of the factorization scale when computing the matched cross section.  In contrast to the NNLL results in Fig.~\ref{mww_14_scales}, the $\mu_f$ dependence of the matched result is less than (similar to) the hard and soft scale dependencies for $\mu<\mu^0$ ($\mu>\mu^0)$.  Also, note that unlike the soft and factorization scales, the hard scale dependence, with a minimum near the central value, actually never decreases but always increases the total cross section as it is varied from the central value.  This explains why in Fig.~\ref{mww_14_scales} the effect of the hard scale variation was always to increase the differential cross section value above the central value.

\subsection{Total Cross Section}

\begin{table}[tp]
\renewcommand{\arraystretch}{1.4}
\caption{  
Total cross sections for $pp\rightarrow W^+W^-$ with $\mu_f^0=2M_{W}$, $\mu_h^0=M_{WW}$, $\mu_s^0=\mu_s^{\rm min}$, and $Q_h^0=M_{WW}$. The NLO $\sigma^{NLO}$ includes the $gg$ contribution and is evaluated with NLO PDFs and the remaining entries are evaluated with MSTW2008nnlo PDFs.  The primed cross section $\sigma^{\prime NLO+NNLL}$ is the sum of $gg$ contribution $\sigma^{gg}$ and the matched $\sigma^{NLO+NNLL}$, while $\sigma^{\prime NNLO}_{approx}$ is the sum of $\sigma^{gg}$ and approximate NNLO, $\sigma^{NNLO}_{approx}$. The last row, $\sigma^{\prime NNLO}$, is our best prediction for the cross section.}
 \centering
 \vskip .2in
\begin{tabular}{|c|c|c|c|c|}
\hline
  $\sigma({\rm pb})$        &$\sqrt{S}=7$~TeV& $\sqrt{S}=8$~TeV & $\sqrt{S}=13$~TeV &$\sqrt{S}$=14~TeV \\ \hline
$\sigma^{NLO}$&$45.7^{+1.5}_{-1.1}$
 &$55.7^{+1.7}_{-1.2}$
 &$110.6^{+2.5}_{-1.6}$
&$122.2^{+2.5}_{-1.8}$
\\
\hline
$\sigma^{gg}$ &$1.0^{+0.3}_{-0.2}$ &$1.3^{+0.4}_{-0.3}$& $3.5^{+0.9}_{-0.7}$&$4.1^{+0.9}_{-0.7}$
\\ \hline 
$\sigma^{NLO+NNLL}$ &$44.9_{-0.6}^{+0.6}$ &$54.8^{+0.7}_{-0.8}$ &$108.2^{+1.3}_{-1.5}$  &$119.5^{+1.5}_{-1.6}$\\\hline
$\sigma^{\prime NLO+NNLL}$&$45.9^{+0.5}_{-0.6}$&$56.1^{+0.7}_{-0.8}$&$111.7^{+1.8}_{-1.6}$&$123.6^{+2.0}_{-1.8}$\\
\hline
$\sigma^{NNLO}_{approx}$ &$45.0^{+0.4}_{-0.1} $ & $54.9^{+0.5}_{-0.05} $ & $108.3_{-0.4}^{+1.0}$ & $119.6_{-0.5}^{+1.2} $ \\ \hline
$\sigma^{\prime NNLO}_{approx}$&$46.0^{+0.4}_{-0.047}$&$56.2^{+0.6}_{-0.1}$&$111.8^{+1.7}_{-1.1}$&$123.7^{+1.8}_{-1.2}$\\
\hline
\end{tabular}
\vspace{-1ex}
\label{sigtab1}
\end{table}

In this section, we compile our final results for the total $W^+W^-$ cross sections at the LHC. In Tables \ref{sigtab1} and \ref{sigtab2}, we show successively improved results for the total cross sections using MSTW2008 PDFs. Both tables fix the central value of $\mu_h^0=M_{WW}$ and $\mu^0_s=\mu_s^{\rm min}$. Table \ref{sigtab1} takes the central factorization scale to be fixed at $\mu_f^0=2M_W$, while Table \ref{sigtab2} uses a dynamical central scale, $\mu_f^0=M_{WW}$. The top line is the NLO result obtained from MCFM~\cite{Campbell:2011bn} (which includes the $gg$ initial state) and is calculated using NLO PDFs. The second line of the tables is the $gg$ contribution, $\sigma^{gg}$,  calculated using MCFM, but with NNLO PDFs (as is appropriate for combining with the NNLL and approximate NNLO results).  The third and fourth rows contain the NNLL matched and approximate NNLO cross sections evaluated with NNLO PDFs but without the $gg$ contribution, $\sigma^{NLO+NNLL}$ and $\sigma^{NLO}_{approx}$, respectively.  The fifth and sixth rows are the same as the third and fourth, but with the $gg$ contribution now included.  The  uncertainties in the matched cross section correspond to taking the central values of the hard, soft, and factorization scales and varying each separately up and down by a factor of $2$.  The uncertainties in the approximate NNLO cross section correspond to varying the factorization and $Q_h$ scales by a factor of two around their central values.  The resulting uncertainties are added in quadrature.  

As noted previously in the discussion of the previous subsection, the factorization scale dependence of the matched cross section is less than that of the NLO cross section. This cancellation is more extreme at $\sqrt{S}=7$ and $8$~TeV.  Hence, even with hard and soft scale variation taken into account the scale dependence of $\sigma^{\prime NLO+NNLL}$ is less than that of $\sigma^{NLO}$.  At $\sqrt{S}=13$ and $14$ TeV, once the uncertainties associated with hard and soft scale variation are taken into account, the scale uncertainty of $\sigma^{\prime NLO+NNLL}$ is similar to or greater than that of $\sigma^{NLO}$.

\begin{table}[tp]
\renewcommand{\arraystretch}{1.4}
\caption{ Total cross sections for $pp\rightarrow W^+W^-$ with $\mu_f^0=M_{WW}$, $\mu_h^0=M_{WW}$, $\mu_s^0=\mu_s^{\rm min}$, and $Q_h^0 = M_{WW}$. The NLO $\sigma^{NLO}$ includes the $gg$ contribution and is evaluated with NLO PDFs and the remaining entries are evaluated with MSTW2008nnlo PDFs.  The primed cross section $\sigma^{\prime NLO+NNLL}$ is the sum of $gg$ contribution $\sigma^{gg}$ and the matched $\sigma^{NLO+NNLL}$, while $\sigma^{\prime NNLO}_{approx}$ is the sum of $\sigma^{gg}$ and approximate NNLO, $\sigma^{NNLO}_{approx}$. The last row, $\sigma^{\prime NNLO}$, is our best prediction for the cross section.}
\centering
 \vskip .2in
\begin{tabular}{|c|c|c|c|c|}
\hline
  $\sigma({\rm pb})$        &$\sqrt{S}=7$~TeV& $\sqrt{S}=8$~TeV & $\sqrt{S}=13$~TeV &$\sqrt{S}$=14~TeV \\ \hline
$\sigma^{NLO}$&$44.8^{+1.2}_{-0.9}$&$54.7^{+1.4}_{-1.0}$&$108.8^{+1.2}_{-1.3}$&$120.3^{+2.0}_{-1.3}$\\
\hline
$\sigma^{gg}$ &$0.9^{+0.2}_{-0.2}$& $1.2^{+0.3}_{-0.1}$& $3.3^{+0.8}_{-0.6}$& $3.7^{+0.7}_{-0.6}$ \\
\hline
$\sigma^{NLO+NNLL}$ 
&$44.7^{+0.5}_{-0.6}$&$54.6^{+0.6}_{-0.8}$&$108.1^{+1.4}_{-1.5}$&$119.4^{+1.6}_{-1.7}$\\
\hline
$\sigma^{\prime NLO+NNLL}$&$45.6^{+0.6}_{-0.6}$& $55.8^{+0.7}_{-0.8}$&$111.4^{+2.0}_{-1.8}$&$123.1^{+2.1}_{-2.0}$\\
\hline
$\sigma^{NNLO}_{approx}$ &$44.8^{+0.4}_{-0.1}$ & $54.7^{+0.4}_{-0.04}$ &$108.2_{-0.4}^{+1.0}$&$119.6_{-0.6}^{+1.2}$ \\ \hline
$\sigma^{\prime NNLO}_{approx}$&$45.7^{+0.4}_{-0.04}$&$55.9^{+0.5}_{-0}$&$111.5^{+1.6}_{-1.0}$&$123.3^{+1.7}_{-1.2}$\\
\hline
\end{tabular}
\vspace{-1ex}
\label{sigtab2}
\end{table}

The scale dependence of the approximate NNLO cross sections, $\sigma^{\prime NNLO}_{approx}$, at $\sqrt{S}=7$ and $8$ TeV
is reduced
 by at least a factor of three relative to the NLO cross section, while at $\sqrt{S}=13$ and $14$ TeV the uncertainties of the NLO and $\sigma^{\prime NNLO}_{approx}$ cross sections are more similar.  This is due to a cancellation of the factorization scale dependence of $\sigma^{gg}$ and $\sigma^{NNLO}_{approx}$ at $\sqrt{S}=7$ and $8$ TeV that is not present 
 at $\sqrt{S}=13$ and $14$ TeV.  This also explains why the scale variation of approximate NNLO cross section without $gg$ contribution is similar to (less than) that with the $gg$ contribution at $7$ and $8$ TeV ($13$ and $14$ TeV.)  Finally, with $\mu^0_f=M_{WW}$ at $\sqrt{S}=8$~TeV, the zero in the scale uncertainty of $\sigma^{\prime NNLO}_{approx}$ indicates that the factor of $2$ of both the factorization and $Q_h$ scales increase the cross section.

Although there is no significant difference between the cross section prediction of $\sigma^{\prime NNLO}_{approx}$ and $\sigma^{\prime NLO+NNLL}$,  we consider $\sigma^{\prime NNLO}_{approx}$ to be our best prediction for the LHC $W^+W^-$ production cross sections, since $\sigma^{\prime NNLO}_{approx}$ for the most part has less scale variation than $\sigma^{\prime NLO+NNLL}$. By comparing the NLO and approximate NNLO, $\sigma^{\prime NNLO}_{approx}$, cross sections, it is apparent that the effect of the higher order corrections is to increase
the cross section less than $\sim 1$~pb at $\sqrt{S}=8$~TeV and less than $\sim 3$~pb at $\sqrt{S}=14$~TeV, while reducing the theoretical uncertainty from scale variations.  The matched NNLL cross section increases the NLO cross section by similar amounts, with a slightly increased scale uncertainty.  There is very little difference between using a fixed factorization scale and a dynamic factorization scale.
It appears that the prediction for the $W^+W^-$ cross section is under good theoretical control.

\section{Conclusion}
\label{conc.sec}
Now that the Higgs boson is discovered, a full exploration of the electroweak sector has begun.  An important signal for this exploration is the pair production of gauge bosons, in particular $W^+W^-$ production.  This signal is a major background to $H\rightarrow W^+W^-$ and is sensitive to the electroweak gauge boson triple coupling, which directly probes the mechanism of electroweak symmetry breaking and the $SU(2)\times U(1)$ gauge structure, respectively.  In order to be sensitive to new physics in the $W^+W^-$ signal and measure the $H\rightarrow W^+W^-$ decay channel well, it is important to have accurate and precise theoretical predictions for $W^+W^-$ production cross section and differential distributions.

In this paper we resummed large logarithms associated with soft gluon emission at partonic threshold, $z=M^2_{WW}/s\rightarrow 1$, at NNLL order for $W^+W^-$ pair production.  This resummation was performed using the formalism of SCET~\cite{Bauer:2000ew,Bauer:2000yr,Bauer:2001yt,Beneke:2002ph} which allows for the resummation directly in momentum space~\cite{Becher:2006mr,Becher:2006nr}.  The NNLL resummed results were then matched onto the known NLO results~\cite{Frixione:1993yp,Ohnemus:1991kk}.  We also calculated the approximate NNLO $W^+W^-$ cross section.  We thus obtain the most accurate cross sections and invariant mass distributions for $W^+W^-$ production that have been calculated to date.

We found that the effect of the threshold resummation on the invariant mass distribution was to increase the differential cross section $\sim 3-4\%$ in the peak region for both $\sqrt{S}=8$ and $14$ TeV.   The matched NLO+NNLL and approximate NNLO cross section both increased the NLO cross section by $\sim 0.5-1.5\%$ for a factorization scale central value $\mu^0_f=2M_W$ and $\sim 2-3\%$ for a central scale of $\mu^0_f = M_{WW}$, within the theoretical uncertainties. The theoretical uncertainties of the approximate NNLO cross section were generally decreased relative to those of the NLO cross section.  These results indicate that the sideband analysis used for $W^+W^-$ background estimation to $H\rightarrow W^+W^-$ signal~\cite{CMS:bxa,ATLAS:2013wla} is not significantly altered by higher order corrections.  Also, the strong coupling constant perturbative expansion of the $W^+W^-$ production cross section is firmly under theoretical control.
\section*{Acknowledgements}
S.D. and I.L. are supported by the U.S. Department of Energy under grant
No.~DE-AC02-98CH10886.  M.Z.  is supported by the National Science Foundation, grant PHY-0969739.
We thank George Sterman and Andrea Ferroglia for helpful conversations. 

\noindent

\appendix
\makeatletter
\renewcommand\section{\@startsection {section}{1}{\z@}%
    {-3.5ex \@plus -1ex \@minus -.2ex}%
    {2.3ex \@plus.2ex}%
    {\normalfont\large\bfseries}}
\makeatother
\renewcommand{\thesubsection}{\Alph{section}.\arabic{subsection}}
\noindent
\section{Fixed Order Results}
\label{appa}
\subsection{Lowest Order Results}
\label{appa:LO}
The coefficients of Eq.~(\ref{lores}) are~\cite{Frixione:1993yp}
\begin{eqnarray}
c_q^{tt}&=&{\pi^2\alpha^2_{EM}\over s_W^2}\nonumber \\
c_q^{ts}(s)&=&{4\pi^2\alpha^2_{EM}\over  s_W^2}
{1\over s}
\biggl(Q_q+{s\over s-M_Z^2}
{1\over s_W^2}(T_{3,q}-Q_qs_W^2)\biggr)\nonumber \\
c_q^{ss}(s)&=&{16\pi^2\alpha^2_{EM}\over s^2}\biggl\{
\biggl(Q_q+{1\over 2 s_W^2}(T_{3,q}-2Q_qs_W^2){s\over s-M_Z^2}\biggr)^2
+\biggl({T_{3,q}\over 2 s_W^2}{s\over s-M_Z^2}\biggr)^2\biggr\}
\end{eqnarray}
with $T_{3,q}=\pm {1\over 2}$ and $s_W=\sin\theta_W$.
The functions occurring in the lowest order amplitudes are,
\begin{eqnarray}
F_u^{0}(s,t)&=&F_d^{0}(s,u)
\nonumber \\
&=&16\biggl(
{ut\over M_W^4}-1
\biggr)
\biggl({1\over 4}+{M_W^4\over t^2}
\biggr)+16{s\over M_W^2}
\nonumber \\
J_u^{0}(s,t)&=&-J_d^{0}(s,u)
\nonumber \\ &=&
16\biggl(
{ut\over M_W^4}-1\biggr)\biggl({s\over 4}-{M_W^2\over 2}-{M_W^4\over t}\biggr)
+16s\biggl({s\over M_W^2}-2+{2M_W^2\over t}\biggr)
\nonumber \\
K_u^{0}(s,t)&=& K_d^{0}(s,u)\nonumber \\
&=&
8\biggl({ut\over M_W^4}-1\biggr)
\biggl({s^2\over 4}-sM_W^2+3M_W^4\biggr)
+8s^2\biggl({s\over M_W^2}-4\biggr)
\, .
\end{eqnarray}
\subsection{NLO Results}
\label{appa:NLO}
The functions occurring in the one-loop virtual amplitude are\cite{Frixione:1993yp},
\begin{eqnarray}
F_u^{1}(s,t)&=&{4(80t^2+73st-140M_W^2t+72M_W^4) \over t^2}
-{4(4t+s)^2\over s\beta^2 t}
-{128(t+2s)\over M_W^2}\nonumber \\&&
+{64(t+s)\over M_W^4}
-\biggl({32(t^2-3st-3M_W^4)\over t^2}+{128s\over t-M_W^2}\biggr)\log\biggl({-t\over M_W^2}\biggr)
\nonumber \\
&&+\biggl({8(6t^2+8st-19M_W^2t+12M_W^4)\over t^2}-{32t^2-128st-26s^2\over s\beta^2t}
+{6(4t+s)^2\over s\beta^4t}\biggr)\log\biggl({s\over M_W^2}\biggr)
\nonumber \\
&&+32s\biggl({2M_W^4\over t}-u\biggr)I_4
-64 (t-M_W^2)\biggl(
{2M_W^4\over t^2}-{u\over t}\biggr)I_{3t}
\nonumber \\
&&+\biggl(
{16t(4M_W^2-u)-49s^2+72M_W^2s-48M_W^4\over 2t}
+{2(8t^2-14st-3s^2)\over \beta^2t}
-{3(4t+s)^2\over 2\beta^4t}\biggr)I_{3l}
\nonumber \\&&
+{32\pi^2\over 3}\biggl({2(t+2s)\over M_W^2}-{3t+2s-4M_W^2\over t}
-{t(t+s)\over M_W^4}\biggr)
\nonumber \\
J_u^{1}(s,t)
&=&
-{128(t^2+2st+2s^2)\over M_W^2}
-{16(t^2-21st-26M_W^2t+34M_W^2s+17M_W^4)\over t}\nonumber\\
&& +{64st(t+s)\over M_W^4}+{32s^2\over t-M_W^2}
\nonumber \\
&&
+\biggl(16(t-5s+2M_W^2)-{48M_W^2(2s+M_W^2)\over t}
+{64s(2t+s)\over t-M_W^2}
-{32s^2t\over (t-M_W^2)^2}\biggr)\log\biggl({-t\over M_W^2}\biggr)
\nonumber \\
&&
+
\biggl(
{16(4t+s)\over\beta^2}
-16(3t-2s)+{48M_W^2(2t-2s-M_W^2)\over t}\biggr)
\log\biggl({s\over M_W^2}\biggr)
\nonumber \\
&&+16s\biggl(t(2s+u)-2M_W^2(2s+M_W^2)\biggr)I_4
+32(t-M_W^2)\biggl({2M_W^2(2s+M_W^2)\over t}-2s-u\biggr)I_{3t}
\nonumber \\ &&
+\biggl(32st-12s^2+32M_W^4-16M_W^2(2t+7s)-{4s(4t+s)\over\beta^2}\biggr)I_{3l}
\nonumber \\
&&+{32\pi^2\over 3}\biggl({2(t^2+2st+2s^2)\over M_W^2}
-{st(t+s)\over M_W^4}-{2M_W^2(2t-2s-M_W^2)\over t}
-t-4s\biggr)
\nonumber \\
K_u^{1}(s,t)&=& 16\biggl\{12t^2+20st-24M_W^2t+17s^2-4M_W^2s+12M_W^4+
{s^2t(t+s)\over M_W^4}
\nonumber \\
 &&
-{2s(2t^2+3st+2s^2)\over M_W^2}\biggr\}(2-{\pi^2\over 3})
\end{eqnarray}
with $F^1_d(s,t)=F^1_u(s,u)$, $J^1_d(s,t)=-J^2_u(s,u)$, and $K^1_d(s,t)=K^1_u(s,u)$.  The integrals are given by
\begin{eqnarray}
I_4&=&{1\over st}\biggl(2\log^2\biggl({-t\over M_W^2}\biggr)-4\log\biggl({M_W^2-t\over M_W^2}\biggr)\log\biggl({-t\over M_W^2}\biggr)
-4{\rm Li}_2\biggl({t\over M_W^2}\biggr)\biggr)
\nonumber \\
I_{3t}&=&{1\over M_W^2-t}\biggl({1\over 2}\log^2\biggl({M_W^2\over s}\biggr)-{1\over 2}\log^2\biggl({-t\over s}\biggr)
-{\pi^2\over 2}\biggr)
\nonumber \\
I_{3l}&=&{1\over s\beta}
\biggl(4{\rm Li}_2\biggl(
{\beta-1\over 1+\beta}\biggr)
+\log^2\biggl({1-\beta\over 1+\beta}\biggr)
+{\pi^2\over 3}\biggr)\, .
\end{eqnarray}
\noindent
\section{Approximate NNLO Results}
\label{appb}

The hard scattering kernel is expanded in a power series:
\begin{eqnarray}
C(z,M,\cos\theta,\mu)&=&C^{0}(z,M,\cos\theta,\mu)+\frac{\alpha_s}{4\pi}C^{1}(z,M,\cos\theta,\mu)
\nonumber \\ &&
+\left(\frac{\alpha_s}{4\pi}\right)^2C^{2}(z,M,\cos\theta,\mu).
\, .
\end{eqnarray}
Similarly, the hard function is expanded in a power series:
\begin{eqnarray}
H(M_{WW},\cos\theta,\mu_f)&=&H^{0}(M_{WW},\cos\theta) +\frac{\alpha_s}{4\pi}H^{1}(M_{WW},\cos\theta,\mu_f)\nonumber\\&&+\left(\frac{\alpha_s}{4\pi}\right)^2H^{2}(M_{WW},\cos\theta,\mu_f)\label{NNLOexact},
\end{eqnarray}
where $H^1=H^1_{reg}+H^1_{extra}$ where $H^1_{reg}$ and $H^1_{extra}$ are defined in Eq.~(\ref{H1}).

The approximate NNLO cross section is found by calculating the scale dependent pieces of the leading singular contribution to $C^{2}$ and adding this contribution to the total NLO cross section.Using the results for the hard and soft functions to NNLO, an approximate formula for the NNLO piece, $C^{2}$, can be determined which includes the leading singular pieces. The result is written as an expansion of $C^{2}$ in ``plus"-functions:
\begin{eqnarray}
C^{2}(z,M,\cos\theta,\mu_f)=\sum^3_{n=0}D^{(n)}\left[\frac{\ln^n(1-z)}{1-z}\right]_+ + R^{(0)}\delta(1-z),
\end{eqnarray}
where
\begin{eqnarray}
D^{(3)}&=&64 H^{0} s^{(2,4)}\\
 D^{(2)}&=&24 H^{0} \left[s^{(2,3)}+4 L_s s^{(2,4)}\right]\\
D^{(1)}&= &8 H^{0}\left[s^{(2,2)}+3 L_s s^{(2,3)}+6\left(L^2_s-\frac{2\pi^2}{3}\right)s^{(2,4)}\right]+8 H^{(1)} s^{(1,2)} \\
D^{(0)}&=&2H^{0}\left[s^{(2,1)}+2L_s s^{(2,2)}+3\left(L^2_s-\frac{2\pi^2}{3}\right)s^{(2,3)}+4\left(L^3_s-2L_s\pi^2+16\zeta_3\right)s^{(2,4)}\right]\\
&&+2 H^{1}\left[s^{(1,1)}+2L_s s^{(1,2)}\right]\nonumber
\end{eqnarray}
and 
\begin{eqnarray}
R^{(0)}&=&
H^{0}\left[
s^{(2,0)}+
L_M s^{(2,1)}+\left(L^2_M-\frac{2\pi^2}{3}\right)s^{(2,2)}
\right.
\nonumber \\ &&
\left.
+\left(L^3_M-2L_M\pi^2+16\zeta_3\right)s^{(2,3)}\right.\\
&&\left.+\left(L^4_M-4L^2_M\pi^2+\frac{4\pi^4}{15}+64L_M\zeta_3\right)s^{(2,4)}\right]
\nonumber \\
&&+H^{1}\left[s^{(1,0)}+L_Ms^{(1,1)}+\left(L^2_M-\frac{2\pi^2}{3}\right)s^{(1,2)}\right]+H^{2}\, .
\end{eqnarray}
The logarithms are defined as
\begin{eqnarray}
L_M&=&\log\biggl({M^2\over \mu_f^2}\biggr)\nonumber \\
L_s&=&\log\biggl({s\over \mu_f^2}\biggr)
\end{eqnarray}

The soft contributions are found from the RG evolution and explicit calculation of the soft function~\cite{Becher:2007ty},
\begin{eqnarray}
s^{(1,0)}&=&\frac{C_F\pi^2}{3}
\nonumber \\
s^{(2,0)}&=&C_F\left[C_F\frac{\pi^4}{18}+C_A\left(\frac{2428}{81}+\frac{67\pi^2}{54}-\frac{\pi^4}{3}-\frac{22}{9}\zeta_3\right)-T_Fn_f\left(\frac{656}{81}+\frac{10\pi^2}{27}-\frac{8}{9}\zeta_3\right)\right]
\nonumber \\
s^{(1,2)}&=&\frac{\Gamma_0}{2}
\nonumber \\
s^{(1,1)} &=& \gamma^s_0
\nonumber \\
s^{(2,4)}&=&\frac{\Gamma^2_0}{8}
\nonumber \\
 s^{(2,3)}&=&\frac{\Gamma_0}{6}(3\gamma^s_0-\beta_0)
 \nonumber \\
 s^{(2,2)}&=&\frac{1}{2}\left(\Gamma_0 s^{(1,0)}+\Gamma_1 +(\gamma^s_0)^2-\beta_0\gamma^s_0\right)
 \nonumber \\
 s^{(2,1)}&=&s^{(1,0)}(\gamma^s_0-\beta_0)+\gamma^s_1,
\end{eqnarray}
where $C_F=4/3$, $C_A=3$, $T_F=1/2$, $n_f=5$, and $\zeta_3$ is a Riemann zeta function.
Expressions for $\Gamma_0,\Gamma_1,\gamma_0^s, \gamma_1^s$ and $\beta_0$ can be found in Ref.~\cite{Becher:2007ty} (where the soft anomalous dimension is written as $\gamma^W$ instead of $\gamma^s$).

Similarly, the hard coefficients can be expanded as a power series in logs,
\begin{eqnarray}
H^{0}(M_{WW},\cos\theta)&=&h^{(0,0)}(M_{WW},\cos\theta) \\
H^{1}(M_{WW},\cos\theta,\mu_f)&=&\sum^2_{n=0}h^{(1,n)}\left(M_{WW},\cos\theta,\frac{M_{WW}}{Q_h}\right)L^n_Q
\nonumber \\
H^{2}(M_{WW},\cos\theta,\mu_f)&=&\sum^4_{n=0}h^{(2,n)}\left(M_{WW},\cos\theta,\frac{M_{WW}}{Q_h}\right)L^n_Q\nonumber\, ,
\end{eqnarray}
and 
\begin{equation}
L\equiv \ln\biggl({Q_h^2\over \mu_f^2}\biggr) \, .
\end{equation}
We have introduced an additional arbitrary scale $Q_h$.
Using the RGEs of the hard function, we can solve for the hard coefficients:
\begin{eqnarray}
h^{(1,2)}&=&-\frac{\Gamma_0}{2}h^{(0,0)}
\nonumber \\
h^{(1,1)}&=&-\left(\gamma^V_0+\Gamma_0\ln\frac{M^2_{WW}}{Q^2_h}\right)h^{(0,0)}
\nonumber \\
h^{(2,4)}&=&~~\frac{\Gamma^2_0}{8}h^{(0,0)}
\nonumber \\
h^{(2,3)}&=&\frac{\Gamma_0}{6}\left[3\gamma^V_0+\beta_0+3\Gamma_0\ln\frac{M^2_{WW}}{Q^2_h}\right]h^{(0,0)}
\nonumber \\
h^{(2,2)}&=&\frac{1}{2}\left[-\Gamma_0 h^{(1,0)}-\Gamma_1 h^{(0,0)}+\left(\gamma^V_0
+\Gamma_0\ln\frac{M^2_{WW}}{Q^2_h}\right)\left(\gamma^V_0+\beta_0+\Gamma_0\ln\frac{M^2_{WW}}{Q^2_h}\right)h^{(0,0)}\right]
\nonumber 
\\
h^{(2,1)}&=&-\left(\gamma^V_0+\beta_0+\Gamma_0\ln\frac{M^2_{WW}}{Q^2_h}\right)h^{(1,0)}-\left(\gamma^V_1
+\Gamma_1\ln\frac{M^2_{WW}}{Q^2_h}\right)h^{(0,0)},
\end{eqnarray}
where the arguments of the hard coefficients have been suppressed. The anomalous dimension of the hard Wilson coefficient $C_V$, $\gamma^V$, can be found in Ref.~\cite{Becher:2007ty}.  The coefficients $h^{(0,0)}$ and $h^{(1,0)}$ can be calculated from the known LO and NLO hard functions given in Eqs.~(\ref{H0}) and (\ref{H1}).  Additionally, since an additional arbitrary scale $Q_h$ was introduced, the $Q_h$ dependence of $h^{(1,0)}$ and $h^{(2,0)}$ can be solved for:
\begin{eqnarray}
h^{(1,0)}&=&\sum^2_{n=0}h^{(1,n)}_{Q_h=M_{WW}}\ln^n\frac{M^2_{WW}}{Q^2_h}\\
h^{(2,0)}&=&\sum^4_{n=0}h^{(2,n)}_{Q_h=M_{WW}}\ln^n\frac{M^2_{WW}}{Q^2_h}\nonumber\, ,
\end{eqnarray}
where the subscript $Q_h=M_{WW}$ indicates the value of $Q_h$ at which the coefficients on the RHS are evaluated at.

Using these coefficients, the NNLO result in Eq.~(\ref{NNLOexact}) is independent of the scale $Q_h$.  However, without a full calculation, it is not possible to know $h^{(2,0)}$.  Since the other NNLO coefficients, $h^{(2,n)}$ for $n=1,2,3$, are independent of $h^{(2,0)}$, then $h^{(2,0)}$ can be set to zero and an approximate NNLO result is obtained.  The purpose of introducing $Q_h$ is now clear, as discussed in the Section~\ref{approxnnlo}.
\bibliography{paper}
\end{document}